\documentclass[]{article}

\usepackage{amsmath}
\usepackage{authblk}
\usepackage[round]{natbib}
\usepackage{graphicx}
\usepackage{hyperref}
\advance\textheight by \topskip
\author[1]{Jasper B. Yang}
\author[2]{Bryan E. Shepherd}
\author[3]{Thomas Lumley}
\author[1]{Pamela A. Shaw}

\affil[1]{Department of Biostatistics, Epidemiology, and Informatics, University of Pennsylvania, Philadelphia, PA, USA
}
\affil[2]{Department of Biostatistics, Vanderbilt University School of Medicine, Vanderbilt University, Nashville, TN, USA}
\affil[3]{Department of Statistics, University of Auckland, New Zealand}
\begin{document}
%% almost as usual
\title{Optimum Allocation for Adaptive Multi-Wave Sampling in \proglang{R}: The \proglang{R} Package \pkg{optimall}}

%%%%%%
%% From JSS
%%%%%
\newcommand\code{\bgroup\@codex}
\def\@codex#1{{\normalfont\ttfamily\hyphenchar\font=-1 #1}\egroup}
\let\proglang=\textsf
\newcommand{\pkg}[1]{{\fontseries{b}\selectfont #1}}

%% an abstract and keywords
\maketitle
\begin{abstract}
   The \proglang{R} package \pkg{optimall} offers a collection of functions that efficiently streamline the design process of sampling in surveys ranging from simple to complex. The package’s main functions allow users to interactively define and adjust strata cut points based on values or quantiles of auxiliary covariates, adaptively calculate the optimum number of samples to allocate to each stratum using Neyman or Wright allocation, and select specific IDs to sample based on a stratified sampling design. Using real-life epidemiological study examples, we demonstrate how \pkg{optimall} facilitates an efficient workflow for the design and implementation of surveys in \proglang{R}. Although tailored towards multi-wave sampling under two- or three-phase designs, the \proglang{R} package \pkg{optimall} may be useful for any sampling survey.
\end{abstract}

%% publication information
%% NOTE: Typically, this can be left commented and will be filled out by the technical editor
%% \Volume{50}
%% \Issue{9}
%% \Month{June}
%% \Year{2012}
%% \Submitdate{2012-06-04}
%% \Acceptdate{2012-06-04}

%% for those who use Sweave please include the following line (with % symbols):
%% need no \usepackage{Sweave.sty}

%% end of declarations %%%%%%%%%%%%%%%%%%%%%%%%%%%%%%%%%%%%%%%%%%%%%%%

%% include your article here, just as usual
%% Note that you should use the \pkg{}, \proglang{} and \code{} commands.

\section{Introduction}

Sample surveys are widely used by epidemiologists, political scientists, government organizations, and many other researchers to gather information about finite populations when complete enumerations of variables of interest are unreasonable. A common objective for these surveys is efficiency, meaning that the estimates generated from the budget-constrained number of samples aim to be as precise and accurate as possible. Maximizing efficiency under stratified sampling designs requires efficiently allocating samples to strata, but this task is rarely straightforward because the true optimal design frequently relies on unknown parameters. Traditional methods for approximating optimal design include proportional allocation and x-optimal allocation, although the sampling designs that these strategies yield are only efficient under specific assumptions \citep{sarndal2003model}.

Studies in epidemiology present an interesting case when certain variables such as laboratory tests are expensive to collect, but other variables such as demographic characteristics can easily be obtained for all or most of the population from electronic health records. In these conditions, a common approach is a two-phase sampling design, where readily available or inexpensive variables are obtained for all sampling units in phase one and used to inform the stratified sampling scheme for the collection of the expensive variables on a subsample in phase two \citep{Neyman1938,tao2020optimal}. Recent studies have demonstrated that even more precise estimates of the true population parameters can be obtained when the second phase is conducted over a series of adaptive waves \citep{McIsaac, Chen}. Under this approach, which is often referred to as adaptive multi-wave sampling, estimates for the parameters specifying the underlying optimal design are updated after each wave and used to guide the design for the following round of expensive data collection. While theoretically appealing, a drawback of adaptive multi-wave sampling is that its implementation is more tedious and complicated compared to sampling in a single wave. In this article, we describe the \proglang{R} \citep{Rsoftware} package \pkg{optimall}, which facilitates the processes of stratification, sample selection, optimum allocation, and workflow organization in surveys ranging from simple to complex.

A number of useful packages have been developed for the analysis of sampling survey data in \proglang{R}, notably the \pkg{survey} package by \citet{survey_package}, but fewer exist that are specifically tailored towards sampling design. Among the few are \pkg{surveyplanning} by \citet{surveyplanning_package}, which offers tools for the early stages of survey design including precision estimation and basic optimum allocation, and  \pkg{stratifyR} by \citet{reddy2020stratifyr}, which uses a dynamic programming technique to estimate optimum strata boundaries and sample sizes for univariate populations under certain distributional assumptions. \pkg{PracTools} by \citet{PracTools_package} provides functions for sample size calculation and variance estimation in multistage sampling designs, and \pkg{sampling} by \citet{sampling_package} allows users to randomly draw samples based on various sampling schemes. These packages are useful for specific tasks under certain survey conditions, but none provide a framework for an entire sampling workflow. Such a framework may be especially useful during the implementation of multi-wave designs, where the iterative process of stratifying, optimally allocating samples, taking samples, and merging preliminary results can become cumbersome and error-prone. The \pkg{optimall} package seeks to fill this role. To the authors' knowledge, this is the first \proglang{R} package created specifically for multi-wave sampling design.

The \pkg{optimall} package offers a collection of functions that are relatively simple at their core, but they efficiently streamline the design process of sampling studies ranging from simple to complex when implemented together. The package is particularly tailored towards multi-wave sampling under two- or three-phase designs but may be useful for any sampling survey. In the following sections we introduce the methodological framework underlying the functions, describe the key features of \pkg{optimall} and how they are each implemented, and finally demonstrate their use through practical examples based on real-life epidemiological studies. 

\section{Methodological background}

\subsection{Multi-phase sampling designs}

Two-phase (more generally, multi-phase) sampling designs were first introduced by \citet{Neyman1938} as a strategy for developing an efficient sampling design when some variables of interest are expensive or difficult to collect, but others can be collected on a large portion of the population at a low cost. Such a scenario is common in modern epidemiological studies when biomarker test results or hand-validated electronic health records can only reasonably be measured for some patients, but other variables such as demographic information can be extracted for the entire population.

Under two-phase designs, which are detailed in \citet{sarndal2003model}, the objective of the first phase is to measure inexpensive variables on a large sample of the population. The results of this phase are then used to guide the sampling design for the next phase by defining strata boundaries or informing models for optimal selection. Assuming that the phase one variables are correlated at least partially with the phase two variables, this approach leads to a more optimal, and thus cost-efficient, sampling design for the collection of expensive variables in phase two.

\subsection{Neyman allocation}

Assuming that the per-unit sampling cost is the same in each stratum and that $S_h$, the standard deviation of the variable of interest in stratum $h$, can be estimated, \citet{Neyman1934} presented the following solution to optimally allocate $n$ samples among $H$ strata:

\begin{equation}
n_h = n \frac{N_hS_h}{\sum_{i=1}^H N_iS_i},
\end{equation}

where $N_h$ is the number of population elements in stratum $h$ and $n_h$ is the number to be sampled from stratum $h$.
This formula, known as Neyman allocation, is widely used as a robust method for minimizing the variance of an estimate for the population mean. Neyman allocation offers the advantage of producing exact sampling fractions that can later be multiplied by $n$ or be taken on their own if $n$ is not known. However, this theoretical advantage may sometimes lead to sub-optimal practical applications because it means that Neyman Allocation is not constrained to integer solutions. When taking a fraction of a sample is not reasonable, study designers are forced to stray from the theory by rounding stratum sample sizes to the nearest integer in ways that may not be exactly optimal and may no longer sum to $n$.

\subsection{Wright's exact optimization algorithm}

To address the issues involved with practical applications of Neyman allocation, \citet{Wright2012} presented an algorithm for optimum sampling allocation based on the equal proportions method of allocating a fixed number of seats in the United States House of Representatives among states. 

Like Neyman, the purpose of Wright's algorithm is to to minimize the variance of the sample mean, $\hat{T}$, from stratified random sampling, which can be written as

\begin{equation}\label{2}
Var(\hat{T}) = \sum_{h=1}^{H}\frac{N_{h}^{2}S_{h}^{2}}{n_{h}} - \sum_{h=1}^{H}\frac{N_{h}^{2}S_{h}^{2}}{N_{h}}.
\end{equation}

When the total sample size, $n$, is fixed, Wright notes that the variance of $\hat{T}$ can be minimized by focusing on

$$\sum_{h=1}^{H}\frac{N^2_{h}S^2_{h}}{n_{h}}$$

because the second sum in (\ref{2}) is a constant based on the known population sizes in each stratum, and both $N_h$ and $S_h$ must always be positive. Noting that $1/n_h$ can be written as a sum, we have

\begin{equation}\label{3}
\sum_{h=1}^{H}\frac{N^2_{h}S^2_{h}}{n_{h}} = 
\sum_{h=1}^{H}N^2_{h}S^2_{h} - 
\sum_{h=1}^{H}\left(\frac{N^2_{h}S^2_{h}}{1\cdot2} +
\frac{N^2_{h}S^2_{h}}{2\cdot3} + \cdots + 
\frac{N^2_{h}S^2_{h}}{(n_h - 1)(n_h)}
\right) .
\end{equation}

Assuming that at least one sample should be allocated to each stratum, Wright points out that the integer sampling allocation among $H$ strata that minimizes $Var(\hat{T})$ is found by maximizing the sum in the second term of (\ref{3}) under the constraint $\sum_{h=1}^{H}n_h = n$. This is the same as maximizing

\begin{equation}\label{4}
\sum_{h=1}^{H}\left(\frac{N_{h}S_{h}}{\sqrt{1\cdot2}} +
\frac{N_{h}S_{h}}{\sqrt{2\cdot3}} + \cdots + 
\frac{N_{h}S_{h}}{\sqrt{(n_h - 1)(n_h)}}
\right) .
\end{equation}

Referring to each term in the sum above as a "priority value", Wright points out that selecting the largest priority values across all $h$ will maximize the sum. In what he calls Algorithm I, each stratum is required to have at least 1 sample, so the $n - H$ largest values of \ref{4} across all strata determine the optimum allocation. Each stratum $h$ starts with an allocation of 1 sample, and it receives an additional sample for every priority value of $h$ that lies in the largest $n-H$ overall priority values.

Under this framework, any sample size constraints can be easily incorporated into the allocation process. For example, each stratum needs at least 2 samples in order to generate an unbiased estimate of $Var(\hat{T})$. Wright incorporates this in his Algorithm II by assigning 2 samples to each stratum and selecting the remaining $n - 2H$ samples based on the largest priority values but now ignoring the largest priority value per stratum, the $\frac{N_hS_h}{\sqrt{1 \cdot 2}}$ term. 

\subsection{Optimum allocation for regression parameters}

The previous sections describe well-known methods for minimizing the variance of an estimate for the population mean of a variable of interest given a fixed sample size. In practice however, research interest often lies in the relationship between covariates and an outcome of interest, which is explored through regression modelling. \citet{Chen} and \citet{amorim2021two} point out that Neyman and Wright optimum allocation strategies are still useful in these cases because regression parameters can be considered as the sum of their influence functions. Thus, they show that the allocation of samples to strata that minimizes the variance of the estimate of the sum of influence functions leads to the optimal sampling design, which can be written as:
\begin{equation}\label{5}
n_h = n\frac{N_h Var(h_h(\beta))^{1/2}}{\sum_{i = 1}^{H}N_i Var(h_i(\beta))^{1/2}},
\end{equation}
where $Var(h_h(\beta))^{1/2}$ is the standard deviation of influence functions in stratum $h$.
\subsection{Optimum allocation in multi-wave sampling}

The optimum allocation algorithms presented by Neyman and Wright are useful in the design of stratified sampling surveys; however, they rely on standard deviation estimates for the variable of interest which are typically unknown at the design stage. One approach to this issue is to approximate the true standard deviations using information from previous surveys or using an inexpensive auxiliary variable strongly correlated with the variable of interest \citep{sarndal2003model}. A second, and more effective, approach proposed by \citet{McIsaac} is to conduct the sampling of the expensive variables over a series of waves, re-calculating the optimum sampling proportion after each new wave of data collection. Under this multi-wave sampling approach, necessary within-strata standard deviation estimates, and thus the estimated optimal sampling proportions, are expected to be closer to their true value as the validated data accumulate. As such, the benefits of adaptive, multi-wave sampling better assure the cost-effective efficiency gains offered by two-phase (and, more generally, multi-phase) sampling designs.

\begin{figure}[h!]
\begin{center}
\includegraphics[width = 10 cm]{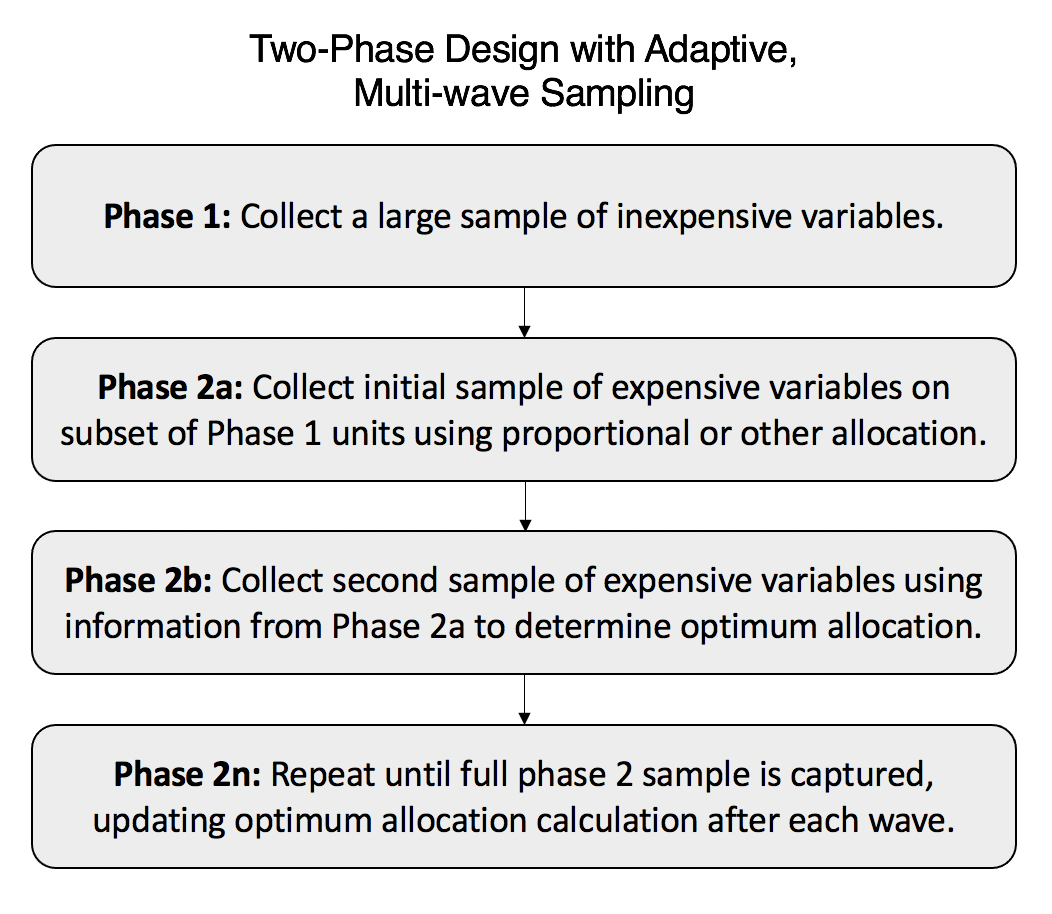}
\caption{Flowchart of multi-wave sampling under a two-phase design.}
\label{Fig1}
\end{center}
\end{figure}

In summary, when full data collection on all subjects is not feasible, we recommend a two-phase sampling design with adaptive, multi-wave sampling in the second phase (Figure \ref{Fig1}). This process is based upon the designs proposed by \citet{McIsaac} and \citet{Chen}. It is applicable across a wide range of epidemiological studies and can be easily implemented in \proglang{R} using the package \pkg{optimall}.  

\section[Features of optimall]{Features of \pkg{optimall}}\label{Section 3}

The \proglang{R} package \pkg{optimall} offers a collection of functions that are designed to streamline the process of optimum sample allocation, specifically under an adaptive, multi-wave approach. Its main functions allow users to
\begin{itemize}
    \setlength\itemsep{.1em}
    \item Define, split, or merge strata based on values or percentiles of variables.
    \item Determine the optimum number or fraction of samples to allocate to each stratum in a given study in order to minimize the variance of a sample mean of interest. 
    \item Select specific IDs to sample based on a given stratified sampling design.
    \item Optimally allocate a fixed number of samples to a sampling wave based on results from a prior wave. 
\end{itemize}

When used together, these functions can automate most of any stratified sampling workflow. This section describes in detail how each of these individual features is implemented, using examples generated with the \code{iris} dataset from the \pkg{datasets} package \citep{Rsoftware}. \pkg{optimall} generally works with data in this format, where each row represents one sampling unit and each column represents one variable.

\subsection{Defining, splitting, and merging strata}

In stratified sampling, strata are typically defined on values or quantiles of variables available for the entire population. Given a dataset with one row per sampling unit and at least one column containing a variable that can be used to construct strata, \pkg{optimall} allows users to easily define, split, and merge strata.

Suppose that we have defined three strata in the \code{iris} dataset based on Species. The dataset contains 50 observations per strata:

\begin{verbatim}
R> table(iris$Species)
 
     setosa versicolor  virginica 
         50         50         50
\end{verbatim}

Suppose that we want to split two out of the three strata, setosa and virginica, based on the within-stratum median of \code{"Sepal.Width"}. In optimall, we can do this by calling the \code{split\char`_strata()} function:
\begin{verbatim}
R> iris2 <- split_strata(data = iris,
+                        strata = "Species",
+                        split = c("setosa", "virginica"), 
+                        split_var = "Sepal.Width",
+                        split_at = c(0.5), type = "local quantile")  

\end{verbatim}

\code{split\char`_strata()} outputs a data frame that matches the input data frame, but contains a new column called \code{"new\char`_strata"} that holds the new strata:
\begin{verbatim}
R> table(iris2$new_strata)
 
  setosa.Sepal.Width_(3.4,4.4]  setosa.Sepal.Width_[2.3,3.4] 
                            22                            28 
                    versicolor virginica.Sepal.Width_(3,3.8] 
                            50                            17 
 virginica.Sepal.Width_[2.2,3] 
                            33
\end{verbatim}
In this example, we split the strata at the within-stratum median, but the \code{split\char`_at} and \code{type} arguments allow us to also perform splits at one or more global quantiles, specific values, or categories for discrete variables as desired. Also note that the new stratum names display the boundaries on which each stratum has been defined. If the new names are too long, the \code{trunc} argument can be used to specify how they should be trimmed.

Similarly, we can merge strata using the \code{merge\char`_strata()} function:
\begin{verbatim}
R> iris3 <- merge_strata(data = iris,
+                        strata = "Species",
+                        merge = c("setosa", "versicolor"), 
+                        name = "set_or_vers")

R> table(iris3$new_strata)
 
 set_or_vers   virginica 
         100          50
\end{verbatim}
\subsection{Optimally allocating samples}

\pkg{optimall} allows users to perform Neyman allocation, Wright Algorithm I, or Wright Algorithm II on any stratified dataset with a continuous variable. The function defaults to using Wright Algorithm II, which requires that at least 2 samples are taken from each stratum. In \pkg{optimall}, stratum sampling sizes for both Wright algorithms are also constrained from above at $N_h$, the population stratum sizes.

Suppose we want to optimally allocate 40 samples among species in the iris dataset, minimizing the variance of the \code{"Sepal.Width"} sample mean. We can use \code{optimum\char`_allocation()}:

\begin{verbatim}
R> sampling_design <- optimum_allocation(data = iris, 
+                                        strata = "Species", 
+                                        y = "Sepal.Width",
+                                        nsample = 40,
+                                        method = "WrightII")
                                         
R> sampling_design

       strata npop   sd  n_sd stratum_fraction stratum_size
 1     setosa   50 0.38 18.95             0.38           15
 2 versicolor   50 0.31 15.69             0.30           12
 3  virginica   50 0.32 16.12             0.32           13
\end{verbatim}

In the example above, \code{optimum\char`_allocation()} determined the optimum sample allocation using standard deviation estimates generated from the full data, which held one row per population unit and contained a single column holding the variable of interest. If we instead have a data frame that holds the $N_h$ and the within-stratum standard deviations, $sd_h$, for each stratum rather than data for each individual unit, we can still use \code{optimum\char`_allocation()}:

\begin{verbatim}
R> iris_summary <- data.frame(strata = unique(iris$Species),
+                             size = c(50, 50, 50),
+                             sd = c(0.3791, 0.3138, 0.3225))

R> optimum_allocation(data = iris_summary, 
+                     strata = "strata",
+                     sd_h = "sd", 
+                     N_h = "size", 
+                     nsample = 40, 
+                     method = "WrightII")
                      
       strata npop   sd  n_sd stratum_fraction stratum_size
 1     setosa   50 0.38 18.95             0.38           15
 2 versicolor   50 0.31 15.69             0.30           12
 3  virginica   50 0.32 16.12             0.32           13
\end{verbatim}

This functionality may be especially useful for more complex optimization problems, such as allocation with respect to multiple parameters.

Regardless of input format, \code{optimum\char`_allocation()} outputs a data frame with one row per stratum, and columns holding the specified total sample size, the stratum fraction, and the number of samples allocated across strata if \code{nsample} is given. We refer to this format as a \emph{design data frame}. Assuming simple random sampling within strata, the design data frame specifies the sampling design for a given wave.
\subsection{Sampling based on results}
With the number of units to sample per stratum specified in a design data frame, \pkg{optimall} can select the IDs of the units to be sampled using simple random sampling within strata with the function \code{sample\char`_strata()}:

\begin{verbatim}
R> iris$id <- 1:150
R> set.seed(743)

R> iris <- sample_strata(data = iris, strata = "Species", id = "id",
+                        design_data = sampling_design, 
+                        design_strata = "strata",
+                        n_allocated = "stratum_size")

\end{verbatim}

The output of \code{sample\char`_strata()} is the same input data frame with a new column called \newline \code{"sample\char`_indicator"} that holds a binary indicator for whether each unit should be sampled in the specified wave:

\begin{verbatim}
R> head(iris)

   Sepal.Length Sepal.Width Petal.Length Petal.Width Species id sample_indicator
 1          5.1         3.5          1.4         0.2  setosa  1                0
 2          4.9         3.0          1.4         0.2  setosa  2                0
 3          4.7         3.2          1.3         0.2  setosa  3                0
 4          4.6         3.1          1.5         0.2  setosa  4                1
 5          5.0         3.6          1.4         0.2  setosa  5                0
 6          5.4         3.9          1.7         0.4  setosa  6                1

\end{verbatim}

From this data frame, we can easily extract a vector of the ids to sample:

\begin{verbatim}
R> ids_to_sample <- iris$id[iris$sample_indicator == 1]
R> head(ids_to_sample)

 [1]  4  6  8 11 14 15
 
R> length(ids_to_sample)

 [1] 40
\end{verbatim}

In cases where some units have already been sampled, \code{sample\char`_strata()} will only select new units if an indicator for previously sampled is specified in the \code{already\char`_sampled} argument. Note that the \code{design\char`_data} argument refers to the output of \code{optimum\char`_allocation()} in this example, but in practice it can be any data frame that has one row per stratum and one column specifying each stratum’s desired sample size. As such, any method for allocating samples to strata can be implemented by \code{sample\char`_strata()} as long as the design data frame is specified and simple random sampling within strata is used.

\subsection{Allocating samples in waves}

In a multi-wave sampling survey similar to the one described in Section 2.5, \pkg{optimall} allows users to determine the optimum allocation for a given wave, taking into consideration the samples from previous waves, with the function \code{allocate\char`_wave()}. This function calculates the optimum sample allocation for the grand total sample size (cumulative of previous wave sizes and current wave size) to minimize the variance of the estimate for a mean of a given continuous variable using the Wright's allocation  algorithm II for a fixed sample size. It then determines the allocation for the current wave by taking the difference between the previous stratum sample size and the total optimum sample size. If some strata have already been oversampled, it re-calculates the optimum allocation among the non-oversampled strata. In this case, the realized allocation will be sub-optimal.

For a simple example, suppose that we again want to allocate 40 samples to minimize the variance of \code{"Sepal.Width"} in the iris dataset, but 30 out of the 40 samples have already been collected in wave 1. We assume that the strata are still defined only by species, and that 16 of the first 30 samples were taken from the virginica species, 7 from setosa, and 7 from versicolor. Assuming we only have those 30 samples to base our within-stratum standard deviation estimates on, we can allocate the next 10 samples using \code{allocate\char`_wave()}.

First, we set up wave 1:

\begin{verbatim}
R> # Collect Sepal.Width from only the 30 samples
R> wave1_design <- data.frame(strata = c("setosa",
+                                        "virginica",
+                                        "versicolor"),
+                             stratum_size = c(7, 16, 7))

R> phase1_data <- subset(datasets::iris, select =  -Sepal.Width)

R> phase1_data$id <- 1:nrow(phase1_data) #Add id column

R> set.seed(234) # for sampling reproducibility
R> phase1_data <- sample_strata(data = phase1_data, 
+                               strata = "Species", id = "id",
+                               design_data = wave1_design,
+                               design_strata = "strata",
+                               n_allocated = "stratum_size")

R> wave1_ids <- iris$id[phase1_data$sample_indicator == 1]

R> wave1_sampled_data <- iris[iris$id %in% wave1_ids, c("id","Sepal.Width")]

R> wave1_data <- merge(phase1_data, wave1_sampled_data, by = "id", 
+                      no.dups =  TRUE, all.x = TRUE)

R> # We have Sepal.Width for our 30 samples
R> table(is.na(wave1_data$Sepal.Width), wave1_data$Species)
        
         setosa versicolor virginica
   FALSE      7          7        16
   TRUE      43         43        34

\end{verbatim}
Then, we can run \code{allocate\char`_wave()} to generate the wave 2 design:

\begin{verbatim}

R> wave2_design <- allocate_wave(data = wave1_data,
+                                strata = "Species", 
+                                y = "Sepal.Width",
+                                already_sampled = "sample_indicator",
+                                nsample = 10, 
+                                detailed = TRUE)

R> wave2_design

       strata npop nsample_optimal nsample_actual nsample_prior n_to_sample   sd
 1     setosa   50              19             15             7           8 0.47
 2 versicolor   50              12              9             7           2 0.29
 3  virginica   50               9             16            16           0 0.24
\end{verbatim}

Notice that in this case, the total optimum sample allocation, \code{"nsample\char`_optimal"}, does not match the realized allocation, \code{"nsample\char`_actual"}, for every stratum in the outputted wave 2 design data frame. This is because we oversampled from the virginica stratum in wave 1, meaning that the optimum stratum sample size among 40 total samples was smaller than the amount of samples already taken in that group. When no oversampling occurs, \code{"nsample\char`_optimal"} will match \code{"nsample\char`_actual"} for every stratum.

Finally, we can use \code{sample\char`_strata} to select specific samples based on this design:

\begin{verbatim}
R> wave2_data <- sample_strata(data = wave1_data, 
+                              strata = "Species", id = "id",
+                              design_data = wave2_design,
+                              design_strata = "strata",
+                              n_allocated = "n_to_sample")
\end{verbatim}

\section{Shiny app for interactively defining strata}

In stratified sampling, efficiency is gained when strata cut points are chosen to minimize within-stratum variances and maximize the variance across strata. In some cases, it is also desirable for each of the strata to have similar sample sizes \citep{amorim2021two}. Determining the split points on which to define strata also typically relies on unknown parameters, but estimates for the within-stratum variances can be obtained through an auxiliary variable correlated with the variable of interest or from previous sampling waves in which the variable of interest was collected. The \code{split\char`_strata()} function in \pkg{optimall} is a versatile tool for defining strata, but it is designed for situations where the strata cut points have already been decided by the user. Running it iteratively to experiment with different splits is possible yet tedious.

To aid with the process of defining strata, \pkg{optimall} offers a shiny app that can be launched with the \code{optimall\char`_shiny()} function. The app allows users to interactively select strata split points and view how the stratum sizes, standard deviations, and optimum allocations react. This section demonstrates the utility of \code{optimall\char`_shiny()}.   

\subsection{Launching the app and uploading data}
\begin{figure}[!hb]
\begin{center}
\fbox{\includegraphics[width = 10 cm]{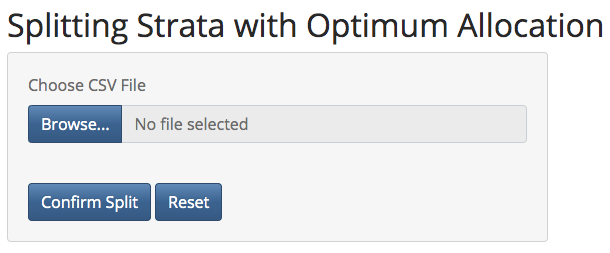}}
\caption{Uploading a csv file in \pkg{optimall}'s shiny application.}
\label{Fig2}
\end{center}
\end{figure}

Launching the app is simple:
\begin{verbatim}
R> optimall_shiny()    
\end{verbatim}

Once open, the first step is uploading data. The data must be in the same format that the rest of the functions in \pkg{optimall} require, meaning that each row must correspond to one sampling unit. As seen in Figure \ref{Fig2}, the data must be saved as a .csv file in order for it to be uploaded in \code{optimall\char`_shiny()}.

\subsection{Reactive strata selection}

After the dataset has been uploaded, options for the user to select inputs will appear. Each input corresponds to an argument of \code{split\char`_strata()}, \code{optimum\char`_allocation()}, or \code{allocate\char`_wave()}. Values for each input can be selected from the list or slider of valid options (Figure \ref{Fig3}):

\begin{figure}[!ht]
\begin{center}
\fbox{\includegraphics[width = 15.2 cm]{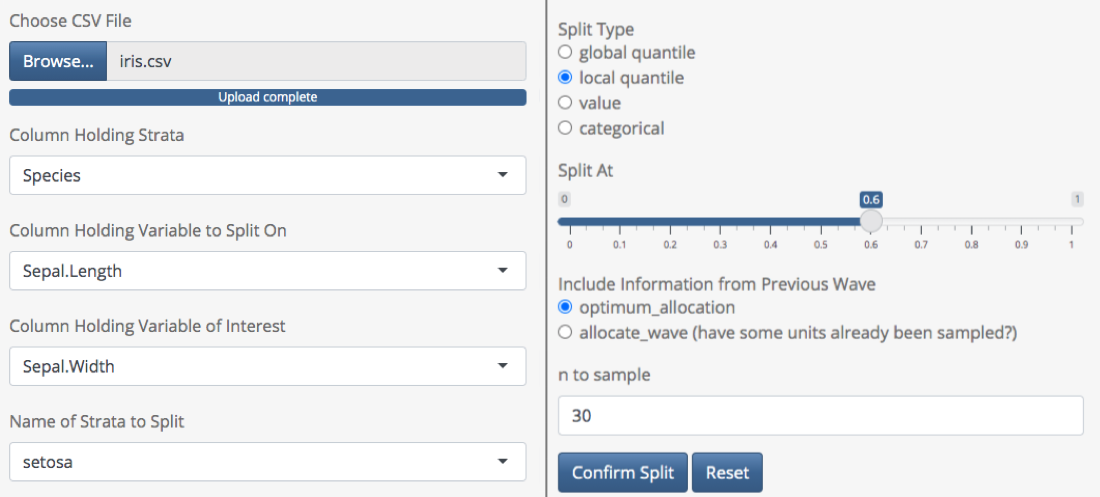}}
\caption{User inputs in the shiny application.}
\label{Fig3}
\end{center}
\end{figure}

In the example shown in Figure \ref{Fig4}, we use the \code{iris} dataset again, splitting the species setosa at the within-stratum 60th percentile of sepal length. The data frame shown in the image below is the design data frame produced by \code{optimum\char`_allocation()}. As we adjust the inputs, this data frame reactively updates to display the within-stratum standard deviations, population sizes, and optimal stratum sample size for our specified parameters. In this way, we can easily observe the impact of different stratum definitions on our sampling design:

\begin{figure}[!ht]
\begin{center}
\fbox{\includegraphics[width = 15 cm]{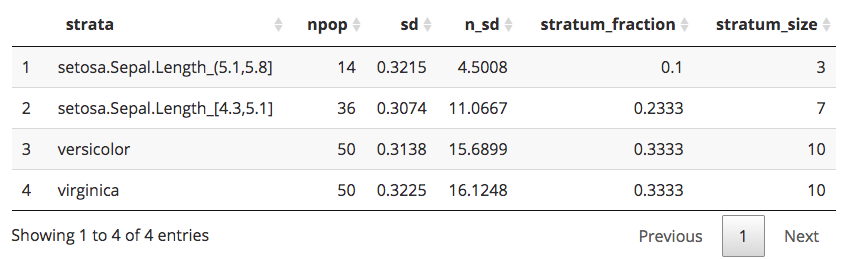}}
\caption{Reactive table output in the shiny application.}
\label{Fig4}
\end{center}
\end{figure}

When we are satisfied with the split parameters, we can select ``Confirm Split" to run \newline \code{split\char`_strata()} with the specified arguments. The displayed data frame will then update to include the new strata, and the code to perform the split in \proglang{R} will appear below it (Figure \ref{Fig5}). Each iteration of this process will add another line of code to the list:

\begin{figure}[!ht]
\begin{center}
\fbox{\includegraphics[width = 15 cm]{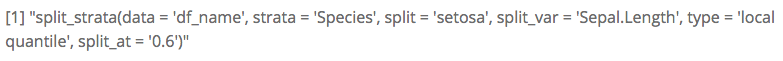}}
\caption{R code output by the shiny application after a split is confirmed.}
\label{Fig5}
\end{center}
\end{figure}

After we have confirmed our desired strata definitions, we can leave the app and perform the splits in our \proglang{R} workflow by copying and pasting the text output of the app. The only change that needs to be added to the pasted code is an updated data frame name. 

\section{The multiwave object}\label{Section 6}

Section \ref{Section 3} demonstrated how the base functions of optimall can be used to efficiently determine optimum allocation, select samples, and split strata during the design of a multi-wave stratified sampling survey. Despite these features, an efficient sampling workflow in R still requires the user to manually organize the many moving parts of the process including design data, a list of samples, data extracted from the samples, and merged data for each wave. When a sampling design involves many waves, these parts can be difficult to keep track of and may be prone to errors. Even more, it may be difficult to go back and reproduce results at the end of a long sampling process.

This section describes an additional feature of optimall called the multiwave object, which stores the metadata, design, samples, and merged data from each step of the multi-wave sampling process in an accessible format. It is optional for the user to use, but it offers the advantages of automatic organization, efficient compatibility with other functions in optimall, and an option for a summary of the sampling design to be printed at any point. The multiwave object contributes towards optimall's goal of being a tool to streamline the often cumbersome aspects of the multi-wave sampling workflow.

\subsection{Format}

\begin{figure}[!hb]
\begin{center}
\includegraphics[width = 12.0 cm]{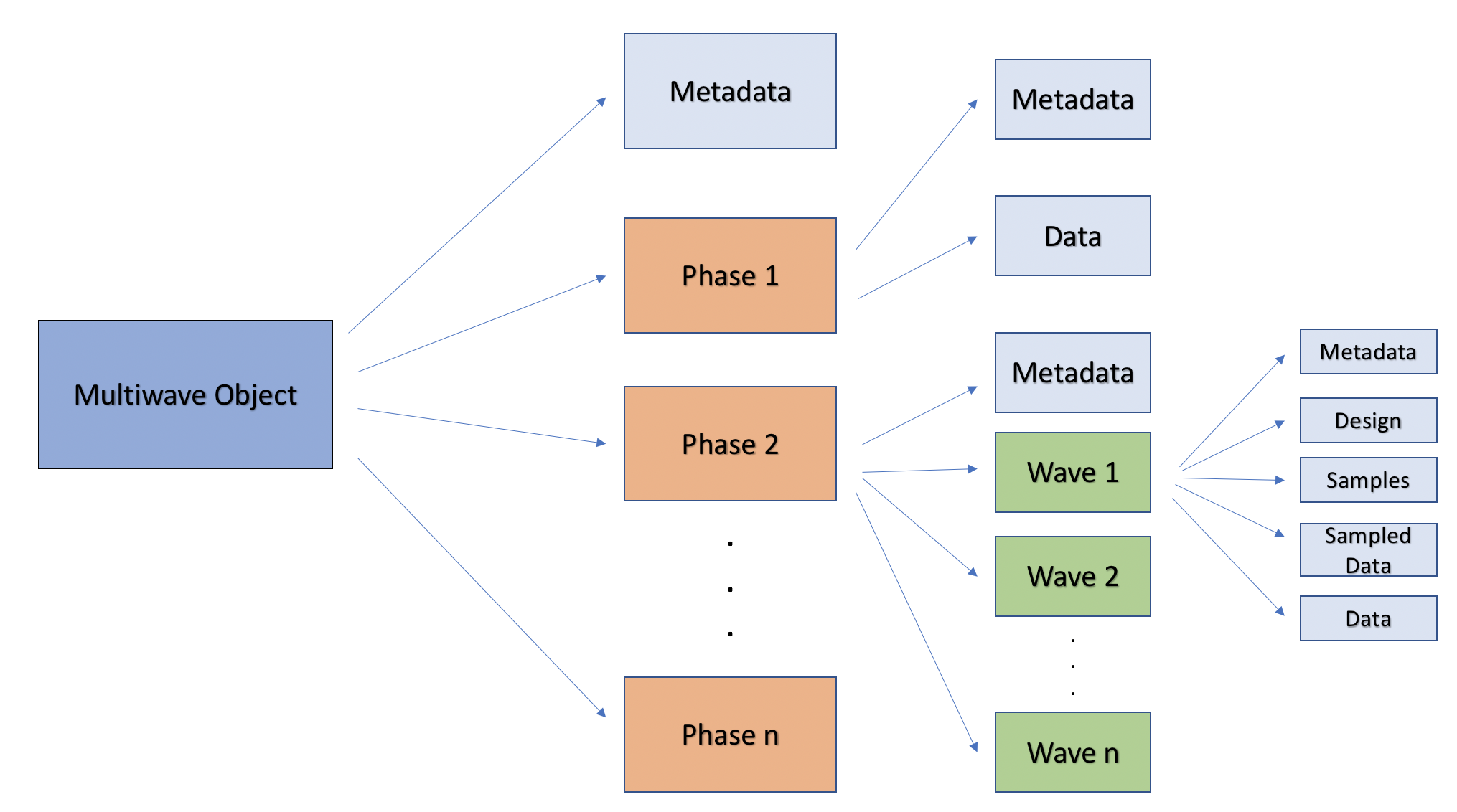}
\caption{Multiwave object structure.}
\label{Fig6}
\end{center}
\end{figure}

The multiwave object uses the S4 class system to store data from an entire multi-wave sampling survey. Its structured is illustrated in Figure \ref{Fig6}. 

The multiwave object contains three S4 classes:

\begin{itemize}
\item \textbf{Multiwave:} The Multiwave class (dark blue in figure above) contains the metadata and a list of Phase objects. It holds the entire sampling design and is the class that the user will interact with the most.
\item \textbf{Phase:} The Phase class (orange in figure above) holds the metadata and a list of Wave objects for each phase.
\item \textbf{Wave:} The Wave class (green in figure above) holds the metadata, design, samples, sampled data, and data for a single wave in multi-wave sampling.

\end{itemize}

The light blue objects in the figure above represent the slots that hold the survey information directly: 
\begin{itemize}
    \item \textbf{Metadata:} The metadata slot holds an initially empty list. Relevant information can be added as named elements, including titles, data dictionaries, and arguments for \pkg{optimall} functions.
    \item \textbf{Design:} The design slot holds the data frame specifying the sampling design of the current wave. In order for it to be used with \code{sample\char`_strata()}, the design data frame must contain at least one column holding the strata names and one column holding the number of samples to be selected from each strata. It is typically the output of \code{optimum\char`_allocation()} or \code{allocate\char`_wave()}.
    \item \textbf{Samples:} The samples slot contains a character vector of the sample ids that were selected in a specific wave.
    \item \textbf{Sampled Data:} The sampled data slot contains a data frame holding the data collected in a specific wave.
    \item \textbf{Data:} The data slot contains a data frame with the full, accumulated data as it gets updated for each phase and wave. It is typically the sampled data merged with the (full) data of the previous wave. In this way, the data in the data slot of a study's ultimate sampling wave is the full study data that can be used for analysis with the \pkg{survey} package. 
\end{itemize}

\subsection{Getting started with the multiwave object}

To work with the multiwave object, users only need to know a few new functions. This section describes how to initialize an object, how to access and write slots of it, and how to deploy some of its useful features. It will generate examples from the \code{iris} dataset, again assuming that sepal width is the initially unknown variable of interest. More advanced examples based on real-life epidemiological studies will be presented in the following section.

A multiwave object is initialized by the function \code{new\char`_multiwave()}:

\begin{verbatim}
R> MySurvey <- new_multiwave(phases = 2, waves = c(1,3))
\end{verbatim}

This code specifies that this survey will use a two-phase design with phase 2 divided into 3 sampling waves. Note that the length of the `waves` argument must match the number of phases. Phase 1 will almost always have one wave.

\subsection{Accessing and writing slots}
As is standard for an S4 object in R, components of the multiwave object are stored in slots. To access and write slots of the multiwave object, the user could use \code{@} and \code{\$}:

\begin{verbatim}
R> # To access overall metadata
R> MySurvey@metadata

list()

R> # To write overall metadata. We may want to include a title.
R> MySurvey@metadata <- list(title = "Sepal Width Survey")

R> # To access Phase 2 metadata
R> MySurvey@phases$phase2@metadata

list()

R> # To access Phase 2, Wave 2 design
R> MySurvey@phases$phase2@waves$wave2@design

Object of class "data.frame"
data frame with 0 columns and 0 rows
\end{verbatim}

but this is overly complicated and potentially unstable. Instead, any slot of the multiwave object can be accessed or written using the function \code{get\char`_data()}:

\begin{verbatim}
R> # To access overall metadata
R> get_data(MySurvey, phase = NA, slot = "metadata")

$title
[1] "Sepal Width Survey"

R> # To write overall metadata
R> get_data(MySurvey, phase = NA, slot = "metadata") <-
+    list(title = "Sepal Width Survey")

R> # To access Phase 2 metadata
R> get_data(MySurvey, phase = 2, slot = "metadata")
 
list()

R> # To access Phase 2, Wave 2 design
R> get_data(MySurvey, phase = 2, wave = 2, slot = "design")

Object of class "data.frame"
data frame with 0 columns and 0 rows

R> # To place iris data in Phase 1
R> get_data(MySurvey, phase = 1, slot = "data") <- 
+    subset(iris, select = -Sepal.Width)

\end{verbatim}
Note that calls to \code{get\char`_data()} from phase 1 do not require a wave to be specified, since phase 1 only consists of one wave.

\subsection{Call functions with fewer arguments}

Another advantage of the multiwave object is that the primary functions of \pkg{optimall} including \newline \code{optimum\char`_allocation()}, \code{allocate\char`_wave()}, and \code{sample\char`_strata()} can be called on the object using the function \code{apply\char`_multiwave()}. The \code{apply\char`_multiwave()} function takes the standard arguments to the function as well as phase and wave, which are used to determine the input data frame(s) and the slot of the object where the output should be placed. If the arguments, including names of columns (which tend to be repetitive when used without the multiwave object framework), are specified in the metadata, the function will find them itself, allowing calls to the function to be much simpler. 

To demonstrate, we return to the example of sampling sepal width from the \code{iris} dataset with an adaptive, multi-wave design. Suppose that we have collected data on sepal length, petal width, and petal length for all 150 iris plants in phase 1, but we have not yet collected any of the ``expensive" sepal width variable. We have already placed the data in the corresponding slot of our multiwave object:

\begin{verbatim}
R> # To place iris data in Phase 1
R> iris <- datasets::iris
R> iris$id <- 1:150
R> get_data(MySurvey, phase = 1, slot = "data") <- 
+    subset(iris, select = -Sepal.Width)  
\end{verbatim}

Now, we want to begin our first wave of sampling sepal width in phase 2. Since we expect that sepal length is correlated with our variable of interest, we decide to x-allocate the first wave of samples using the Wright II algorithm to determine the exact optimum allocation for a fixed sample size, using the already sampled sepal length variable as the design variable. Since we are working in the multiwave object framework, we can use \code{apply\char`_multiwave()} to apply the \code{optimum\char`_allocation()} function:

\begin{verbatim}
R> MySurvey <- apply_multiwave(MySurvey, phase = 2, wave = 1, 
+                              fun = "optimum_allocation", 
+                              strata = "Species", y = "Sepal.Length", 
+                              nsample = 30, method = "WrightII")
\end{verbatim}

Since \code{"strata"} will be \code{"Species"} for every wave, we may instead move that argument to the phase metadata so that we don't have to repetitively specify the same argument in every function call: 

\begin{verbatim}
R> get_data(MySurvey, phase = 2, slot = "metadata") <-
+    list(strata = "Species")
\end{verbatim}

Now we no longer have to specify \code{strata} in the function call:

\begin{verbatim}
R> MySurvey <- apply_multiwave(MySurvey, phase = 2, wave = 1, 
+                              fun = "optimum_allocation", 
+                              y = "Sepal.Length", 
+                              nsample = 30, method = "WrightII")
\end{verbatim}

In the absence of a specific \code{strata} argument, \code{apply\char`_multiwave()} turns to the wave, phase, and then overall metadata to look for the missing argument. In this case, it finds \code{strata = "Species"} in the phase metadata. By specifying the \code{phase} and \code{wave} in the function call, we are telling \code{optimum\char`_allocation()} to use the full data from the prior wave as input and to output the results in the corresponding slot of the specified wave. As such, both calls to \code{apply\char`_multiwave()} output an updated multiwave object with the results of \code{optimum\char`_allocation()} in the phase 2, wave 1 design slot:

\begin{verbatim}
R> get_data(MySurvey, phase = 2, wave = 1, slot = "design")
      strata npop   sd  n_sd stratum_fraction stratum_size
1     setosa   50 0.35 17.62             0.23            7
2 versicolor   50 0.52 25.81             0.33           10
3  virginica   50 0.64 31.79             0.43           13

\end{verbatim}

The \code{allocate\char`_wave()} function can be applied to multiwave objects in the same manner and will be demonstrated in the following examples section. After these functions have been used to specify a "design" data frame (or a manually created design data frame has been placed in the design slot to implement a different allocation strategy), we can use \code{apply\char`_multiwave()} to apply \code{sample\char`_strata()} and select the ids to sample for the given sampling wave:

\begin{verbatim}
R> set.seed(340) # for sampling reproducibility
R> MySurvey <- apply_multiwave(MySurvey, phase = 2, wave = 1, 
+                              fun = "sample_strata", id = "id",
+                              design_strata = "strata",
+                              n_allocated = "stratum_size")    
\end{verbatim}

Note that we did not have to specify the \code{data} or \code{design\char`_data} as we do in the standard version of \code{split\char`_strata()} because they are extracted using the \code{phase} and \code{wave} arguments. We also did not have to specify the \code{strata} argument again because it was available in the phase metadata. The result of this call to \code{apply\char`_multiwave()} is an updated \code{MySurvey} with a character vector of ids to sample in the samples slot:

\begin{verbatim}
R> get_data(MySurvey, phase = 2, wave = 1, slot = "samples")
 [1] "1"   "4"   "22"  "26"  "28"  "30"  "39"  "51"  "53"  "66"  "70"  "73"  
[13] "79"  "80"  "88"  "95"  "99"  "101" "111" "119" "120" "121" "130" "138"
[25] "139" "140" "143" "144" "148" "150"
\end{verbatim}

When working with a multiwave object, a new function called \code{merge\char`_samples()} also becomes available. This function allows users to quickly and efficiently merge the sampled data with the accumulated data from the previous phases and waves. Suppose that we have collected the sepal width for these 30 plant ids and placed the data in the sampled data slot of phase 2, wave 1 of \code{MySurvey}: 

\begin{verbatim}
R> get_data(MySurvey, phase = 2, wave = 1, slot = "sampled_data") <-
+    iris[iris$id %in% get_data(MySurvey, 
+                               phase = 2, 
+                               wave = 1, 
+                               slot = "samples"),
+         c("id", "Sepal.Width")]
\end{verbatim}

We can call \code{merge\char`_samples()} to smoothly merge the sampled data of the current wave with the previously accumulated data (in this case, only the phase 1 data) and place the output in the data slot of the current wave:

\begin{verbatim}
R> MySurvey <- merge_samples(MySurvey, phase = 2, wave = 1,
+                            id = "id", sampled_ind = "sampled_phase2")
\end{verbatim}

In the data slot of phase 2, wave 1, we will now have an updated dataframe with all of the phase 1 data and a sepal width column that has the sampled data for the selected ids and \code{NA} values for the rest of the plants. There is also a new column called \code{"sampled\char`_phase2"} that holds an indicator for which samples have been sampled in phase 2 thus far:

\begin{verbatim}
> head(get_data(MySurvey, phase = 2, wave = 1, slot = "data"))

  id Sepal.Length Petal.Length Petal.Width Sepal.Width sampled_phase1
1  1          5.1          1.4         0.2         3.5              1
2  2          4.9          1.4         0.2          NA              0
3  3          4.7          1.3         0.2          NA              0
4  4          4.6          1.5         0.2         3.1              1
5  5          5.0          1.4         0.2          NA              0
6  6          5.4          1.7         0.4          NA              0
\end{verbatim}

Calls to \code{merge\char`_samples()} in later waves of phase 2 will update the phase sampled indicator each time. The \code{sample\char`_strata()} function will use this column to ensure that units sampled in previous waves are not selected again. The utility of \code{merge\char`_samples()} and \code{apply\char`_multiwave()} are demonstrated further in Section \ref{Section 6}.

\subsection{View summary diagram of survey}

At any point during the multi-wave sampling workflow, \pkg{optimall} allows users to view a diagram of the structure of their survey with \code{multiwave\char`_diagram()}.  This function produces an graphic illustrating the structure using the package \pkg{DiagrammeR} by \citet{diagrammer}. A simpler version can also be printed to the console using \code{summary()}. In the output below we see that the title of the survey, “Sepal Width Survey”, is found from the overall survey metadata and that boxes are colored depending on whether they have been filled yet. Filled slots are blue and contain a short description of their contents. This function enables users to track their progress during a multiwave sampling survey:

\begin{verbatim}
R> multiwave_diagram(MySurvey) 
\end{verbatim}

\begin{center}
\includegraphics[width = 11 cm]{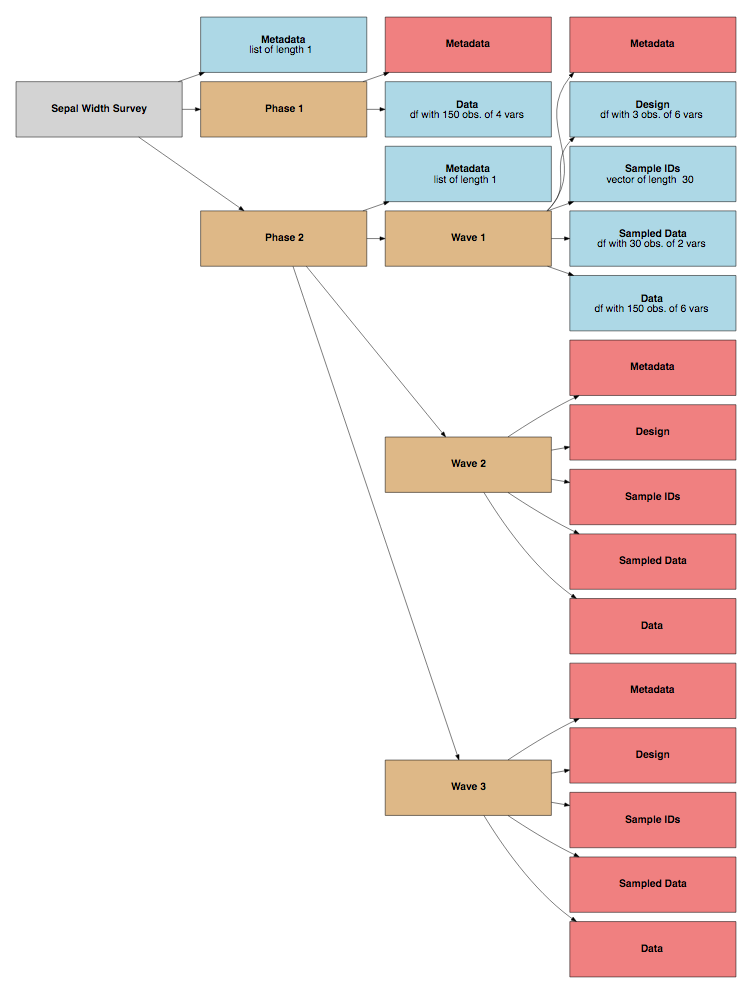}
\end{center}

\section{Examples}

\subsection{Introduction and data set up}

In this section, we use the simulated dataset \code{MatWgt\char`_Sim} to demonstrate how the functions of \pkg{optimall} can be used together to carry out the design of a sampling survey in \proglang{R}. This dataset is based on a real epidemiological study and includes measures of maternal weight gain during pregnancy, diabetes status, and race. The study illustrates another example of when multi-phase, multi-wave sampling is useful: data validation. Suppose that we have access to an inexpensive but error-prone measure of maternal weight change during pregnancy, but we are interested in the true weight change that can only be obtained through expensive data validation. In phase 1, we collect error-prone maternal weight change, race, and diabetes status on all 10,335 patients in our target population. In phase 2, we can afford to validate 750 samples. We can use the adaptive, multi-wave workflow to efficiently allocate these validations.

The raw dataset is available in \pkg{optimall} and is organized as follows:

\begin{verbatim}
R> head(MatWgt_Sim)
    id mat_weight_est  race diabetes obesity mat_weight_true
1 7667      12.954711 Asian        0       0       10.768935
2 8554       9.607124 Asian        0       1        8.892049
3 6324      10.998284 Asian        0       0       12.075861
4 8320      10.702210 Asian        0       0       10.887584
5 8543       6.485988 Asian        1       0        7.331542
6  127      13.338445 Asian        0       1       12.936826
\end{verbatim}

In our example, \code{mat\char`_weight\char`_true} is expensive to collect and unknown initially, so our phase 1 data excludes it:

\begin{verbatim}
R> phase1 <- subset(MatWgt_Sim, select = -mat_weight_true)
\end{verbatim}

The following subsections demonstrate how \pkg{optimall} facilitates the sampling of the true maternal weight change under different conditions.
\subsection{Estimating the Mean of an expensive covariate in a single wave}

Suppose that we are interested in estimating the population mean of the true maternal weight change with minimal variance, but we have to allocate all 750 validation samples in a single wave. Stratified random sampling will be useful for efficiency in this scenario, and we can use the error-prone maternal weight change measure to approximate the optimum allocation for the true measure. To implement this in \pkg{optimall}, we first define the strata. We will split the population into non-overlapping strata using race and global percentiles of maternal weight gain:

\begin{verbatim}
R> # Initialize strata column
R> phase1$strata <- phase1$race

R> # Merge the smallest race categories into "other"
R> phase1 <- merge_strata(data = phase1, 
+                         strata = "strata",
+                         merge = c("Other","Asian"),
+                         name = "Other")

\end{verbatim}

Recall that \pkg{optimall} places the updated strata defined by \code{merge\char`_strata()} and \newline \code{split\char`_strata()} in a new column called \code{"new\char`_strata"}. We next want to split this new column based on quantiles of estimated maternal weight change:

\begin{verbatim}
R> # Split each race stratum at 25th and 75th global percentile of wgt change
R> phase1 <- split_strata(data = phase1, strata = "new_strata", split = NULL, 
+                         split_var = "mat_weight_est", 
+                         type = "global quantile", 
+                         split_at = c(0.25,0.75),
+                         trunc = "MWC_est")
\end{verbatim}

Which gives us our final strata:

\begin{verbatim}
R> table(phase1$new_strata) # 9 strata

 Black.MWC_est_(15.06,38.46]  Black.MWC_est_(9.75,15.06] 
                         628                        1154 
 Black.MWC_est_[-30.21,9.75] Other.MWC_est_(15.06,30.94] 
                         745                         325 
  Other.MWC_est_(9.75,15.06]  Other.MWC_est_[-5.39,9.75] 
                         929                         456 
 White.MWC_est_(15.06,51.69]  White.MWC_est_(9.75,15.06] 
                        1631                        3084 
 White.MWC_est_[-25.68,9.75] 
                        1383
    
\end{verbatim}

With the strata and phase 1 data defined, we can initialize a multiwave object to store the information from this survey:

\begin{verbatim}
R> Survey1 <- new_multiwave(phases = 2, waves = c(1,1))

R> # Add a title
R> get_data(Survey1, phase = NA, slot = "metadata") <-
+    list(title = "Single Wave Mat Wgt Survey")

R> # Add Phase 1 data
R> get_data(Survey1, phase = 1, slot = "data") <- phase1
\end{verbatim}

We can now use \code{apply\char`_multiwave()} to generate the design for the single wave in phase 2. We have not yet sampled any values of the true maternal weight change (our variable of interest), but we do have access to the error-prone maternal weight change variable from phase 1 for every patient in our study. The true and error-prone versions are likely highly correlated, so we can approximate the optimum allocation of samples by applying Neyamn (or Wright) allocation to the error-prone version. This approach of is often referred to as x-optimal allocation \citep{sarndal2003model}. 

\begin{verbatim}
R> Survey1 <- apply_multiwave(Survey1, phase = 2, wave = 1, 
+                             fun = "optimum_allocation",
+                             strata = "new_strata",
+                             y = "mat_weight_est",
+                             nsample = 750,
+                             method = "WrightII")  

R> get_data(Survey1, phase = 2, wave = 1, slot = "design")
                       strata npop   sd    n_sd stratum_fraction stratum_size
1 Black.MWC_est_(15.06,38.46]  628 3.63 2277.53             0.09           68
2  Black.MWC_est_(9.75,15.06] 1154 1.46 1688.46             0.07           51
3 Black.MWC_est_[-30.21,9.75]  745 4.09 3050.57             0.12           92
4 Other.MWC_est_(15.06,30.94]  325 2.41  781.93             0.03           24
5  Other.MWC_est_(9.75,15.06]  929 1.44 1334.35             0.05           40
6  Other.MWC_est_[-5.39,9.75]  456 2.46 1121.03             0.05           34
7 White.MWC_est_(15.06,51.69] 1631 3.51 5726.61             0.23          172
8  White.MWC_est_(9.75,15.06] 3084 1.44 4453.92             0.18          134
9 White.MWC_est_[-25.68,9.75] 1383 3.24 4485.44             0.18          135
\end{verbatim}

We can then use this design and the \code{apply\char`_multiwave()} function to select the specific ids to sample:

\begin{verbatim}
R> # Set seed for reproducibility
R> set.seed(435) 

R> # Select samples
R> Survey1 <- apply_multiwave(Survey1, phase = 2, wave = 1,
+                             fun = "sample_strata",
+                             strata = "new_strata",
+                             id = "id",
+                             design_strata = "strata",
+                             n_allocated = "stratum_size")
\end{verbatim}

We can simulate the data validation process by selecting the true maternal weight changes for these ids and placing the resulting data frame in the sampled data slot:

\begin{verbatim}
R> get_data(Survey1, phase = 2, wave = 1, slot = "sampled_data") <-
+    MatWgt_Sim[MatWgt_Sim$id %in% get_data(Survey1, 
+                                           phase = 2,
+                                           wave = 1,
+                                           slot = "samples"),
+               c("id","mat_weight_true")]
\end{verbatim}

Finally, we can merge the sampled data with the phase 1 data using \code{merge\char`_samples()}:

\begin{verbatim}
R> Survey1 <- merge_samples(Survey1, phase = 2, wave = 1,
+                           id = "id", sampled_ind = "sampled_phase2")
\end{verbatim}

We have now completed validation of 750 samples! The data slot of the final wave (phase 2, wave 1 in this case) holds a data frame that can easily be used to generate estimates with the \pkg{survey} \citep{survey_package} package:

\begin{verbatim}
R> library("survey")

R> # Extract final data
R> final_data <- get_data(Survey1, phase = 2, wave = 1, slot = "data")

R> # Specify two-phase design
R> twophasedesign <- twophase(id = list(~1,~1),
+                             strata = list(NULL,~new_strata),
+                             subset = ~sampled_phase2,
+                             data = final_data, method = "approx")

R> # Stratified Mean Estimate
R> survey::svymean(~mat_weight_true, design = twophasedesign)

                  mean     SE
mat_weight_true 12.363 0.1003

\end{verbatim}

\subsection{Estimating the mean of an expensive covariate over multiple waves}\label{Section 6.3}

For this next example, we again suppose that we can afford to validate the true maternal weight change of 750 patients, but this time we will do it over a series of three adaptive waves. The phase 1 data is unchanged from the previous example, so initializing a multiwave object to capture this survey is straightforward:

\begin{verbatim}
R> Survey2 <- new_multiwave(phases = 2, waves = c(1,3))

R> # Add a title
R> get_data(Survey2, phase = NA, slot = "metadata") <-
+    list(title = "Multi-wave Mat Wgt Survey")

R> # Add Phase 1 data and metadata
R> get_data(Survey2, phase = 1, slot = "data") <- phase1

R> get_data(Survey2, phase = 1, slot = "metadata") <-
+    list(title = "Phase 1",
+         description = "This is simulated data 
+         collected in December 2020 based on a real maternal
+         weight study",
+         data_dict = data.frame(id = "unique identifier",
+                                new_strata = "Name of updated strata",
+                                race = "self-identified race of mother",
+                                mat_weight_est = "error-prone estimate of 
+                                maternal weight change in lbs. during 
+                                pregnancy",
+                                obesity = "1/0 indicator for child obesity 
+                                within 6 years of birth",
+                                diabetes = "1/0 indicator for mother's 
+                                diabetes status"
                                 ))
\end{verbatim}

We will now demonstrate the second phase of sampling by breaking it into waves.

\subsubsection{Phase 2, wave 1}

In the first wave of phase 2, we will allocate 250 samples using proportional allocation. While \code{optimum\char`_allocation()} will not create a proportional allocation design data frame for us, we can easy implement other allocation strategies by manually creating the design data frame:

\begin{verbatim}
R> get_data(Survey2, phase = 2, wave = 1, slot = "design") <- 
+    data.frame(
+      strata_name = names(table(phase1$new_strata)), 
+      strata_n = round(250.3*as.vector(table(phase1$new_strata))/10335)
+      ) # 250.3 to make sure 250 samples after rounding

R> get_data(Survey2, phase = 2, wave = 1, slot = "design")

                  strata_name strata_n
1 Black.MWC_est_(15.06,38.46]       15
2  Black.MWC_est_(9.75,15.06]       28
3 Black.MWC_est_[-30.21,9.75]       18
4 Other.MWC_est_(15.06,30.94]        8
5  Other.MWC_est_(9.75,15.06]       22
6  Other.MWC_est_[-5.39,9.75]       11
7 White.MWC_est_(15.06,51.69]       40
8  White.MWC_est_(9.75,15.06]       75
9 White.MWC_est_[-25.68,9.75]       33
\end{verbatim}

Before we begin the sampling process, we can specify function arguments and other details in the metadata of phase 2:

\begin{verbatim}
R> get_data(Survey2, phase = 2, slot = "metadata") <-
+     list(title = "Phase 2",
+          description = "This phase will sample 750 patients across three
+          waves of 250. The first wave will use proportional allocation,
+          and the next two will use Wright Algorithm II based on the
+          previous wave's samples",
+          strata = "new_strata",
+          id = "id",
+          y = "mat_weight_true")
\end{verbatim}

We can now select the 250 wave 1 ids with \code{sample\char`_strata()}:

\begin{verbatim}
R> set.seed(983) # for sampling reproducibility
R> Survey2 <- apply_multiwave(Survey2, phase = 2, wave = 1, 
+                             fun = "sample_strata",
+                             design_strata  = "strata_name",
+                             n_allocated = "strata_n")
\end{verbatim}

Then we can ``collect" the sampled data:

\begin{verbatim}
R> get_data(Survey2, phase = 2, wave = 1, slot = "sampled_data") <-
+    MatWgt_Sim[MatWgt_Sim$id %in% get_data(Survey2, 
+                                           phase = 2,
+                                           wave = 1,
+                                           slot = "samples"),
+               c("id","mat_weight_true")]
\end{verbatim}

and then merge it with the phase 1 data to get an updated dataset in the data slot using \code{merge\char`_samples()}:

\begin{verbatim}
R> Survey2 <- merge_samples(Survey2, phase = 2, wave = 1,
+                           sampled_ind = "sampled_phase2")

\end{verbatim}

With that, wave 1 is complete!

\begin{verbatim}
R> dim(get_data(Survey2, phase = 2, wave = 1, slot = "data"))

[1] 10335    9

R> table(is.na(get_data(Survey2, phase = 2, wave = 1, "data")$mat_weight_true))

FALSE  TRUE 
  250 10085
\end{verbatim}

\subsubsection{Phase 2, wave 2}

For the next wave, we essentially repeat the process of wave 1. This time, though, we use the 250 samples from wave 1 to estimate the optimal design with \code{allocate\char`_wave()}. Recall from Section 3.4 that \code{allocate\char`_wave()} will output a data frame holding the optimum allocation based on the previous samples of \code{y} using Wright algorithm II. When used within \code{apply\char`_multiwave}, the output design data frame will be automatically placed in the design slot of the specified wave:

\begin{verbatim}
R> Survey2 <- apply_multiwave(Survey2, phase = 2, wave = 2, 
+                             fun = "allocate_wave",
+                             already_sampled = "sampled_phase2",
+                             nsample = 250)
\end{verbatim}

Looking at the design, we see that the proportional allocation in wave 1 did not appear to oversample from any strata:

\begin{verbatim}
R> get_data(Survey2, phase = 2, wave = 2, slot = "design")
                       strata npop nsample_actual nsample_prior n_to_sample
1 Black.MWC_est_(15.06,38.46]  628             35            15          20
2  Black.MWC_est_(9.75,15.06] 1154             36            28           8
3 Black.MWC_est_[-30.21,9.75]  745             72            18          54
4 Other.MWC_est_(15.06,30.94]  325             14             8           6
5  Other.MWC_est_(9.75,15.06]  929             30            22           8
6  Other.MWC_est_[-5.39,9.75]  456             26            11          15
7 White.MWC_est_(15.06,51.69] 1631            111            40          71
8  White.MWC_est_(9.75,15.06] 3084            100            75          25
9 White.MWC_est_[-25.68,9.75] 1383             76            33          43
\end{verbatim}

We use this design to inform the selection of ids to sample:

\begin{verbatim}
R> set.seed(584) # for sampling reproducibility
R> Survey2 <- apply_multiwave(Survey2, phase = 2, wave = 2,
+                             fun = "sample_strata",
+                             already_sampled = "sampled_phase2",
+                             design_strata  = "strata",
+                             n_allocated = "n_to_sample")
\end{verbatim}

then we ``collect" the data from these samples:

\begin{verbatim}
R> get_data(Survey2, phase = 2, wave = 2, slot = "sampled_data") <-
+    MatWgt_Sim[MatWgt_Sim$id %in% get_data(Survey2, 
+                                           phase = 2,
+                                           wave = 2,
+                                           slot = "samples"),
+               c("id","mat_weight_true")]
\end{verbatim}

and merge it back with the data from the previous wave to get the updated data in the data slot of wave 2:

\begin{verbatim}
R> Survey2 <- merge_samples(Survey2, phase = 2, wave = 2,
+                           sampled_ind = "sampled_phase2")

\end{verbatim}

With that, wave 2 is complete!

\begin{verbatim}
R> dim(get_data(Survey2, phase = 2, wave = 2, slot = "data"))

[1] 10335    9

R> table(get_data(Survey2, phase = 2, wave = 2, "data")$sampled_phase2)

   0    1 
9835  500 
\end{verbatim}

\subsubsection{Phase 2, wave 3}
We proceed with wave 3 in the exact same manner as in wave 2. This subsection leaves out some of the specific details covered in the previous waves for brevity.

Allocate the samples:

\begin{verbatim}
R> Survey2 <- apply_multiwave(Survey2, phase = 2, wave = 3, 
+                             fun = "allocate_wave",
+                             already_sampled = "sampled_phase2",
+                             nsample = 250)

\end{verbatim}

Select the samples:

\begin{verbatim}
R> set.seed(584) # for sampling reproducibility
R> Survey2 <- apply_multiwave(Survey2, phase = 2, wave = 3,
+                             fun = "sample_strata",
+                             already_sampled = "sampled_phase2",
+                             design_strata  = "strata",
+                             n_allocated = "n_to_sample")

\end{verbatim}

``Collect" the data:

\begin{verbatim}
R> get_data(Survey2, phase = 2, wave = 3, slot = "sampled_data") <-
+    MatWgt_Sim[MatWgt_Sim$id %in% get_data(Survey2, 
+                                           phase = 2,
+                                           wave = 3,
+                                           slot = "samples"),
+               c("id","mat_weight_true")]

\end{verbatim}

Merge the sampled data:

\begin{verbatim}

R> Survey2 <- merge_samples(Survey2, phase = 2, wave = 3,
+                           sampled_ind = "sampled_phase2")

\end{verbatim}

We have now collected all 750 samples. We can use the data in phase 2, wave 3 to find the mean with the \pkg{survey} package using the same steps as in the previous section.

\subsection{A multi-wave sampling design for regression modelling}

Suppose that we have the same phase 1 dataset for the maternal weight change study, but we are now interested in modelling the relationship between the odds of childhood obesity and true maternal weight change during pregnancy, with diabetes and race as covariates. We will consider childhood obesity as the binary outcome variable and regress it on the covariates using logistic regression. This problem closely resembles those described in \citet{Chen} and \citet{McIsaac}. In this example, we will conduct the sampling over two waves, with 250 samples in the first wave and 500 in the second:

\begin{verbatim}
R> Survey3 <- new_multiwave(phases = 2, waves = c(1,2))

R> get_data(Survey3, phase = NA, slot = "metadata") <- 
+    list(title = "Survey for Regression of Childhood Obesity")

R> get_data(Survey3, phase = 1, slot = "data") <- phase1
\end{verbatim}

\subsubsection{Phase 2, wave 1}

Optimum allocation for a regression covariate is slightly different from the simple mean estimation from the previous examples. In the two-phase sampling setting, estimation of a weighted regression model from the complete case phase 2 data may become more efficient by making use of auxiliary phase 1 variables. The technique, known as survey calibration or raking, uses an auxiliary variable available on the whole cohort to adjust (calibrate) the usual Horwitz-Thompson weights \citep{breslow2009improved,deville1992calibration}. This approach can achieve the optimum minimum variance in the class of ``augmented" inverse probability weighted estimates in the case of the ideal auxiliary variable, which is the expected value of the target parameter’s influence function given the observed data \citep{lumley2011connections,robins1994estimation}. The plug-in approach described in \citet{kulich2004improving} involves approximating the ideal raking variable by first imputing the unobserved phase 2 variables on those not sampled in phase 2 sample and then obtaining the influence function for the target parameter from the model fit with the imputed data. This approach was used in the setting of case-cohort designs in \citet{breslow2009using} and in case-matching in \citet{rivera2016using}. Others suggest repeatedly estimating this influence function with multiple imputation to estimate the expected value of the influence function given the observed data \citep{han2019combining,han2016}. If correctly modeled, this approach will achieve maximal efficiency.

For wave 1 of this example, we employ a simplified version of the Kulich and Lin approach, which uses the influence function from the naive logistic regression model fit with the inexpensive, error-prone phase 1 data for maternal weight to approximate the desired optimal design to estimate this regression coefficient. We can use \code{apply\char`_multiwave()} to apply \code{optimum\char`_allocation()} to the estimated influence functions to allocate the first 250 samples:

\begin{verbatim}
R> # Model fit using error-prone maternal weight
R> fit <- glm(obesity ~ mat_weight_est + diabetes + race, 
+             data = get_data(Survey3, phase = 1, slot = "data"), 
+             family = "quasibinomial")

R> # Fisher information 
R> design_mat <- model.matrix(fit)
R> I_hat <- (t(design_mat) %*% (design_mat * 
+                                fit$fitted.values * 
+                                (1 - fit$fitted.values))) / nrow(design_mat)

R> # Calculate influence function
R> influence <- (design_mat * resid(fit, type = "response")) %*% solve(I_hat)
R> influence <- as.data.frame(influence)

R> # Add it as new column to phase 1 data
R> get_data(Survey3, phase = 1, slot = "data")$naive_influence <-
+    influence[,"mat_weight_est"]
\end{verbatim}

Note that we used the Fisher information matrix and the generalized linear model residuals to estimate the influence function for our parameter of interest. As discussed by \citet{breslow2009using}, influence functions of the Cox model can be readily approximated with the delta-beta residuals in the Cox model fit using the \code{coxph()} function in the \pkg{survey} R package \citep{Therneau2020-xf}. We can now generate the wave 1 design with \code{optimum\char`_allocation}:

\begin{verbatim}
R> Survey3 <- apply_multiwave(Survey3, phase = 2, wave = 1, 
+                             fun = "optimum_allocation",
+                             strata = "new_strata",
+                             y = "naive_influence",
+                             nsample = 250,
+                             method = "WrightII") 
\end{verbatim}

The rest of Phase 2 proceeds as demonstrated in Section \ref{Section 6.3}, only now with the influence function as the variable of interest. We thus leave out the details describing the process in \pkg{optimall} to avoid repetition.

\section{Summary}
This article has demonstrated the utility of the \pkg{optimall} package in stratified surveys ranging from simple to complex. For simple stratified surveys, \code{split\char`_strata()}, \code{merge\char`_strata()}, and the shiny app launched by \code{optimall\char`_shiny()} are straightforward tools for making and implementing stratification decisions, while \code{sample\char`_strata()} is useful for sample selection when simple random sampling is used within strata. For complex multi-phase and multi-wave survey designs, the multiwave object framework streamlines the previously cumbersome implementation process. The functions \code{optimum\char`_allocation()} and \code{allocate\char`_wave()} facilitate Neyman and Wright allocation, which are broadly applicable, but the framework supports the manual implementation of other allocation strategies. 

The overarching goal of \pkg{optimall} is to facilitate a streamlined design process for stratified sampling surveys in \pkg{R}, including those that employ multi-phase and multi-wave designs. While the current version closely aligns with this goal, user feedback will help guide improvements over time. We encourage these comments to be provided at the package's Github page (\href{https://github.com/yangjasp/optimall}{https:://github.com/yangjasp/optimall}). For now, future directions include adding functions to implement other methods for optimum allocation and increasing compatibility with the \pkg{survey} package.

\section*{Acknowledgements} We would like to thank Gustavo Amorim, PhD for providing helpful comments on the paper and package vignettes. This work was supported in part by the U.S. National Institutes of Health (NIH) grant R01-AI131771 and Patient
Centered Outcomes Research Institute (PCORI) Award R-1609-36207. The statements in this manuscript are solely
the responsibility of the authors and do not necessarily represent the views of PCORI or NIH. 
%% Comment out when not using bibtex
%\bibliography{ref}

\bibliographystyle{plainnat}  
%%% comment out the ``thebibliography'' section if using the .bib file.

\end{document}